\documentclass[3p,times,twocolumn]{elsarticle}
\usepackage{fancyhdr}
\usepackage{algpseudocode}
\usepackage{algorithm}
\usepackage[T1]{fontenc}
\usepackage{graphicx}
\usepackage{epstopdf}
\usepackage{amsmath}
\usepackage{subfigure}
\usepackage{etoolbox}
\usepackage{booktabs}
\usepackage{mdframed}
\usepackage{placeins}
\definecolor{darkgreen}{rgb}{0,0.5,0}
\biboptions{numbers,sort&compress}

\usepackage{setspace}

\newcommand{\mbf}[1]{\mathbf{#1}}

\newcommand{\avg}[1]{\ensuremath{\langle #1 \rangle}} 
\newcommand{\pd}[2]{\frac{\partial #1}{\partial #2}}

\newcommand{\Ray}{\ensuremath{\mathit{Ra}}} 
\newcommand{\Pra}{\ensuremath{\mathit{Pr}}} 
\newcommand{\Nus}{\ensuremath{\mathit{Nu}}} 

\journal{}

\author[mms]{Gijs L. Kooij}
\author[macs,mos]{Mikhail A. Botchev}
\author[mms]{Edo M.A. Frederix}
\author[mms,tue]{Bernard J. Geurts\corref{cor1} (\texttt{b.j.geurts@utwente.nl})} \ead{b.j.geurts@utwente.nl}
\author[imp,ucla]{Susanne Horn}
\author[pof,mpi]{Detlef Lohse}
\author[pof]{Erwin P. van der Poel}
\author[mpi]{Olga Shishkina}
\author[pof]{Richard J.A.M. Stevens}
\author[tor,pof]{Roberto Verzicco}
\cortext[cor1]{Corresponding author}

\address[mms]{Multiscale Modeling and Simulation, Faculty EEMCS, University of Twente, Enschede, Netherlands}
\address[imp]{Applied Mathematics and Mathematical Physics, Imperial College London}
\address[pof]{Physics of Fluids, Faculty of Science and Technology, University of Twente, Enschede, Netherlands}
\address[macs]{Mathematics of Computational Science, Faculty EEMCS, University of Twente, Enschede, Netherlands}
\address[mos]{Keldysh Institute of Applied Mathematics, Russian Academy of Sciences, Moscow, Russia}
\address[mpi]{Max Planck institute for Dynamics and Self-Organization, G\"ottingen, Germany}
\address[tor]{Dipartimento di Ingegneria Meccanica, Universit\`a di Roma ``Tor Vergata'', Rome, Italy}
\address[tue]{Faculty of Applied Physics, Fluid Dynamics Laboratory, Eindhoven University of Technology, Eindhoven, Netherlands}
\address[ucla]{Earth, Planetary, and Space Sciences, University of California, Los Angeles, USA}

\title{Comparison of computational codes for direct numerical simulations of turbulent Rayleigh--B\'enard convection}

\begin{document}
\begin{abstract}
Computational codes for direct numerical simulations of Rayleigh-B\'enard (RB) convection are compared in terms of computational cost and quality of the solution. 
As a benchmark case, RB convection at $\Ray = 10^8$ and $\Pra = 1$ in a periodic domain, in cubic and cylindrical containers is considered. A dedicated second-order finite-difference code (\textsc{AFID}/\textsc{RBflow}) and a specialized fourth-order finite-volume code (\textsc{Goldfish}) are compared with a general purpose finite-volume approach (\textsc{OpenFOAM}) and a general purpose spectral-element code (\textsc{Nek5000}). Reassuringly, all codes provide predictions of the average heat transfer that converge to the same values. The computational costs, however, are found to differ considerably. The specialized codes \textsc{AFID}/\textsc{RBflow} and \textsc{Goldfish} are found to excel in efficiency, outperforming the general purpose flow solvers \textsc{Nek5000} and \textsc{OpenFOAM} by an order of magnitude with an error on the Nusselt number $\Nus$ below $5\%$. However, we find that $\Nus$ alone is not sufficient to assess the quality of the numerical results: in fact, instantaneous snapshots of the temperature field from a near wall region obtained for deliberately under-resolved simulations using \textsc{Nek5000} clearly indicate inadequate flow resolution even when $\Nus$ is converged. Overall, dedicated special purpose codes for RB convection are found to be more efficient than general purpose codes.
\end{abstract}
\begin{keyword}
Direct Numerical Simulations, Rayleigh-B\'enard convection, heat transfer.
\end{keyword}
\maketitle

\section{Introduction}
\noindent Rayleigh-B\'enard (RB) convection is the flow driven by buoyancy forces when a fluid layer is heated from below and cooled from above \cite{kad01,ahl09,loh10,chi12}. The main governing parameter for RB convection is the Rayleigh number $\Ray$ which is the ratio between the destabilizing buoyancy and the stabilizing viscous and diffusive effects. For sufficiently large $\Ray$, RB flow becomes turbulent. To understand the flow physics, direct numerical simulation (DNS) is in principle a straightforward approach that can be used to study turbulence dynamics and heat transfer. For low $\Ray$ this is now routinely done. For higher $\Ray$ however, it is much more challenging, though there are many scientific questions. For example, how does the heat transfer scale with the $\Ray$ number in the regime of very high $\Ray$ numbers ($\Ray \gtrsim 10^{14}$)? This regime, referred to as `ultimate', is characterized by an enhanced heat transfer and associated with a transition to fully turbulent boundary layers \cite{gro00,gro11,he12}. DNS, provided it achieves proper accuracy, could be very useful in providing a deeper insight into the nature of this transition and the properties of this ultimate state. Unfortunately, DNS becomes exceedingly demanding as $\Ray$ increases, since turbulence produces smaller flow scales that need finer spatial resolution and proportionally small time steps to track their dynamics. For $\Ray$ where the ultimate regime is expected the needed computational power is, at present, prohibitive and understanding which numerical code is most cost efficient is a key issue to establish a roadmap for the computer simulations of turbulent RB convection.

Over the years, several codes suitable for DNS of turbulent RB convection have been developed. We compare four of these codes. The first is based on the work by Verzicco {\it et al.}\ \cite{ver96,ver03} in which a second-order energy conserving finite-difference method is applied. For the periodic domain simulation we use the \textsc{AFID} code, which was developed by Van der Poel {\it et al.}\ \cite{poe15cf}. For the cylindrical simulations we use the latest version of \textsc{RBflow}, which is an optimized version of the code used by Stevens {\it et al.}\ \cite{ste10,ste11}. The second code is \textsc{Goldfish} by Shishkina {\it et al.}\ \cite{shi15,shi16b,shi16c}, which is based on a finite-volume approach and uses discretization schemes of fourth-order in space. \textsc{Goldfish} can be used to study turbulent thermal convection in cylindrical and parallelepiped domains. The third code is a general purpose open-source code \textsc{Nek5000}, based on the spectral-element method described by Fischer \cite{fis97}. This code is designed to handle a large variety of flow problems, and was also used in the context of RB convection \cite{sch13,koo15}. The fourth code is an open-source software package \textsc{OpenFOAM} \cite{wel98}. More precisely, its widely used second-order finite-volume scheme was selected for the comparison.

In this study, we compare the four codes in terms of computational efficiency and quality of the results, with a special focus on the heat transport by the turbulent flow, measured by the Nusselt number ($\Nus$). The efficiency of the codes is assessed in relation to their computational costs and the capability to achieve grid converged results. We simulate RB convection in three different geometries, i.e., a periodic domain, a cubic container, and a cylindrical container. Experiments for RB convection are typically conducted in a cylindrical tank. A major challenge to the DNS is to handle sharp gradients in the boundary layers near the walls as well as to capture thermal structures (plumes) that protrude far into the bulk of the flow.

In this paper we perform a convergence test at $\Ray=10^8$ and $\Pra=1$ in which we compare several levels of mesh refinement. We show that all four codes produce the same $\Nus$ number when appropriate spatial resolution is used. At high resolutions, the results become practically identical, taking into account a small uncertainty due to the finite averaging time. When we increase the $\Ray$ number for a fixed spatial resolution, we observe, not surprisingly, that for all codes eventually the resolution becomes insufficient for accurately resolving the turbulent flow. However, the $\Nus$ number calculation seems to be more robust against deliberate underresolution in the higher order codes like \textsc{Nek5000} than in the lower order codes. In this context, \textsc{Nek5000} follows the theoretical scaling of $\Nus$ versus $\Ray$ better than the others. This might suggest that, for a given number of grid points, the \textsc{Nek5000} code is capable of correctly capturing the flow physics even when the other codes fail. A direct inspection of some instantaneous snapshots of temperature in the near wall region, however, clearly shows that this is not the case since the temperature distribution displays the footprint of the underlying discretization. The conclusion is that the evaluation of the $\Nus$ alone is not a sufficient criterion to assess the quality of the results that, instead, should be assessed by evaluating more than one quantity. In this paper we also discuss some other advantages and drawbacks of the compared codes.

The remainder of this paper is organized as follows. In \textsection \ref{sec:governing}, we describe the governing equations of RB convection and the geometries of the domains included in this study. The codes are described in more detail in \textsection \ref{sec:num_methods} and the results are compared in \textsection \ref{sec:comparison}. Last, a summary and conclusions are given in \textsection \ref{sec:conclusion}.

\section{Governing equations and evaluation of the Nusselt number} \label{sec:governing}
\noindent In this section we present the mathematical model and introduce the methods adopted to evaluate the $\Nus$ number. We consider RB convection in three different geometries: a periodic domain, a cube, and a cylinder. Every considered RB cell is characterized by a width $D$ and a height $H$. A flow in any RB cell is determined by the dimensionless parameters, which are the Rayleigh number $\Ray = g\beta \Delta H^3/(\nu \kappa)$, the Prandtl number $\Pra = \nu / \kappa$, and the aspect-ratio $\Gamma = D/H$. Here $g$ is gravitational acceleration, $\beta$ the thermal expansion coefficient, $\Delta$ the temperature difference between the upper and lower plate, $\nu$ the kinematic viscosity, and $\kappa$ the thermal diffusivity of the fluid. In this study, we consider $\Gamma =1$ for all geometries, and $\Pra=1$, which means that the inner length scales of the velocity and the temperature fields are of similar order.

In the numerical simulations, we solve the incompressible Navier-Stokes equation with the Boussinesq approximation to account for buoyancy effects. The governing equations read
\begin{align}
\pd{\mbf{u}}{t} + \mbf{u} \cdot \nabla \mbf{u} &= \sqrt{\frac{\Pra}{\Ray}} \nabla^2 \mbf{u} - \nabla p + \theta \mbf{e}_z, \\
\nabla \cdot \mbf{u} &= 0, \\
\pd{\theta}{t} + \mbf{u} \cdot \nabla \theta &= \frac{1}{\sqrt{\Pra \Ray}} \nabla^2 \theta,
\end{align}
where $\mbf{u}$ is the velocity, $p$ the pressure, $\theta$ the temperature, and $\mbf{e}_z$ the unit vector in the vertical direction anti-parallel to the gravitational acceleration. Here lengths are expressed in terms of $H$, velocities in terms of free fall velocity $U=\sqrt{\beta g \Delta H}$, and temperatures in terms of $\Delta$. No-slip and constant temperature conditions are imposed at the plates, and no-slip and adiabatic conditions at the sidewall. We do not simulate all geometrical configurations with all codes, since not every geometry is feasible in every code. For the periodic domain, we compare \textsc{AFID} and \textsc{Nek5000}, and for the cubic container \textsc{Goldfish} and \textsc{Nek5000}. The cylindrical domain is simulated with all four codes included in this study.

One of the main aspects of RB convection is the heat transported by the turbulent flow from the lower to the upper plate. The heat transfer is quantified by the dimensionless heat flux, i.e. the $\Nus$ number which is the ratio of the actual specific heat flux to the purely conductive counterpart. Following \cite{ste10} we consider several ways to compute $\Nus$. First we consider those of them, which are related directly to the gradient of the temperature and to the convective heat transport. As a function of the vertical coordinate z, $\Nus$ is defined as the average heat flux through a horizontal cross section of the domain \cite{ver99},
$Nu(z) = -\langle\partial_z \theta\rangle_A + \sqrt{Pr\,Ra} \langle u_z \theta\rangle_A$
where $\avg{\cdot}_A$ denotes the average over a horizontal cross section $A$ and in time. From the no-slip boundary conditions, it follows that the $\Nus$ numbers at the lower and upper plate, denoted by $\Nus_{lo}$ and $\Nus_{up}$ respectively, can be calculated from the average temperature gradient at the plates only. We also define the average of the two as $\Nus_{pl} := (\Nus_{lo} + \Nus_{up})/2$. The third definition is obtained using the volume average $\Nus_{vol} := 1 + \sqrt{\Ray \Pr} \avg{u_z \theta}_V$, where $\avg{\cdot}_V$ denotes the average over the complete volume of the domain. Note that these definitions of $\Nus$ are averaged over time as well.

Two more definitions can be obtained from the global balance of energy. We can derive a relation between the $\Nus$ number and the kinetic and thermal dissipation rates \cite{shr90}. The kinetic and thermal dissipation rates are
$\varepsilon := \sqrt{\Pra/\Ray} (\nabla \mbf{u})^2$,
$\varepsilon_\theta := 1/(\sqrt{\Pra \Ray}) (\nabla \theta)^2$ 
and the $\Nus$ number can be calculated from the kinetic and thermal dissipation rate respectively as follows: %
$\Nus_{kin} := 1 + \sqrt{\Pra \Ray}\,\avg{\varepsilon}_V$, and \label{eq:nu_kin}
$\Nus_{th} := \sqrt{\Pra \Ray}\,\avg{\varepsilon_\theta}_V$. \label{eq:nu_th}
These relations from the global balance of energy are sometimes used to assess the quality of DNS of RB convection. If the simulation is well resolved, the global balance of energy is respected accurately. When averaged over time $\Nus$ calculated from the dissipation rates agrees with the other definitions of the $\Nus$ number. Note that the converse is not necessarily true as will be illustrated in \textsection \ref{sec:convergence}. In particular, if some definitions of $\Nus$ agree with each other then this does not automatically imply that the resolution is adequate.

Being an integral quantity, $\Nus$ is one of the main characteristics in RB convection and is, therefore, a natural quantity to investigate. Of course, besides $\Nus$, there are other quantities describing the turbulent RB flow that one could include in a comparison. A good prediction of $\Nus$ does not automatically guarantee that other quantities are approximated accurately, in particular higher order moments will converge less easily. However, the converse holds, i.e., $\Nus$ predictions will correspond closely if the solution is accurately captured. In the present comparison study, we focus mostly on $\Nus$, because it is one of the most important quantities and it gives a first indication of how well different codes perform.

\section{Numerical methods}
\label{sec:num_methods}
\noindent In this section, we provide a brief description of the four codes that are compared. Detailed information can be found in the mentioned references.

\subsection{\textsc{AFID}/\textsc{RBflow}}
\noindent The second-order finite-difference scheme has initially been developed by Verzicco {\it et al.}\ \cite{ver96,ver03} for cylindrical containers. Time integration is performed with a third-order Runge-Kutta method, in combination with a second-order Crank-Nicolson scheme for the viscous terms. \textsc{RBflow}, which is used for the simulations in the cylindrical domain, computes all viscous terms implicitly. The open-source code \textsc{AFID}, specialized for domains with two periodic horizontal directions, uses an explicit scheme in the non-bounded directions to improve scalability of the code \cite{poe15cf}. In \textsc{AFID}, the pressure is solved using a fast Fourier Transform (FFT) in the horizontal directions by means of a 2D pencil decomposition \cite{poe15cf}.

\subsection{\textsc{Nek5000}}
\noindent The open-source package \textsc{Nek5000} is based on the spectral element method, which is an essential extension of the standard finite element method to the case of higher-order basis functions. In this case, the basis functions for the velocity and the pressure are tensor product Lagrange polynomials of order $\mathcal{N}$. Details of the code are found in Ref.\ \cite{fis97}. The spectral element method has been used successfully for DNS of RB convection \cite{koo15,sch13}. We use the so-called $P_\mathcal{N}$-$P_\mathcal{N}$ formulation, based on the splitting scheme in Ref.\ \cite{tom97}. This means that the spectral elements for the velocity components and the pressure are both order $\mathcal{N}$. In the more traditional $P_\mathcal{N}$-$P_{\mathcal{N}-2}$ formulation the pressure is treated with order $\mathcal{N}-2$. Slightly more accurate results can be obtained with the higher order approximation of the pressure in the $P_\mathcal{N}$-$P_\mathcal{N}$ formulation for moderately resolved turbulent flows. In our simulations, we use $N = 8$, which is similar to what was used in DNS of other turbulent flows in \cite{sch13,elk13,ohl10}. The viscous term is treated implicitly with the second-order backward differentiation formula, in combination with an explicit second-order extrapolation scheme for the convective and other terms. The linear system for the velocity is solved with the conjugate gradient method using Jacobi preconditioning. The linear system for the pressure is solved with the generalized minimal residual method, preconditioned with an additive Schwarz method.

\subsection{\textsc{Goldfish}}
\noindent The computational code \textsc{Goldfish} is based on a finite-volume approach. To calculate the velocity and temperature at the surfaces of each finite volume, it uses higher-order discretization schemes in space, up to the fourth order in the case of equidistant meshes. \textsc{Goldfish} has been used to study thermal convective flows in different configurations \cite{shi16c,shi15,shi16b}, in cylindrical and parallelepiped domains. For the time integration, the leapfrog scheme is used for the convective term and the explicit Euler scheme for the viscous term. Although formally first order in time, the accuracy is close to second-order in convection dominated flows \cite[Section 5.8]{wes01}. Note that due to the von Neumann numerical stability of the chosen scheme, the fourth-order spatial discretization requires asymptotically at least 4/3 times finer time stepping than the second-order scheme \cite{shi07b}. Due to the regularity of the used computational meshes, direct solvers are applied to compute the pressure in cylindrical and Cartesian coordinate systems. Thus, when the RB container is a cylinder, FFT is used in two directions. In the case of a parallelepiped RB container, the grid regularity also allows separation of variables. In this case, the corresponding eigenvalues and eigenvectors for the pressure solver are calculated and stored at the beginning of the simulations. The code is quite flexible in parallelization, including parallel I/O, and is characterized by high modularity and is applicable to different configurations of turbulent thermal convective flows.

\subsection{\textsc{OpenFOAM}}
\noindent \textsc{OpenFOAM} is a widely used open-source second-order finite-volume software package \cite{wel98}. Although \textsc{OpenFOAM} offers many different options, we use the implementation of a standard solver in \textsc{OpenFOAM}, which would be representative for typical engineering applications. A linear interpolation scheme is used for the convective term. The equations are solved with the PISO algorithm. The default implementation of the second-order Crank-Nicolson scheme is used for time integration. Furthermore, we do not use the ``non-orthogonal'' correction for the non-orthogonality of the mesh.

\section{Performance comparison}
\label{sec:comparison}
\noindent In this section, we present the results of the simulations using two specialized RB convection codes (\textsc{AFID}/\textsc{RBflow} and \textsc{Goldfish}) and two general purpose codes (\textsc{Nek5000} and \textsc{OpenFOAM}). Our findings shed some light on the issue of the relevance of general purpose codes for moderate $\Ray$ number turbulence. We first discuss a convergence test for a moderate $\Ray=10^8$, for which a fully resolved DNS is easily affordable. A comparison in terms of quality of results and cost is made in \textsection \ref{sec:convergence}. Finally, some results at higher $\Ray$ numbers and fixed spatial resolution are shown in \textsection \ref{sec:high_ra}, illustrating the inevitable loss of accuracy with significant rise of $\Ray$.

The simulations with \textsc{AFID/RBflow}, \textsc{Nek5000}, and \textsc{OpenFOAM}, are all performed on Cartesius (SURFsara). The simulations with \textsc{Goldfish} are performed on SuperMUC of the Leibniz-Rechenzentrum (LRZ). All the simulations run on Cartesius are performed on the same type of `thin' nodes with Intel Haswell processors clocked at 2.6 Ghz. Similar nodes were also used on SuperMUC. Hence, we expect similar performance on these computing platforms.

\begin{figure*}[!t]
\begin{center}
\includegraphics[width=0.33\textwidth]{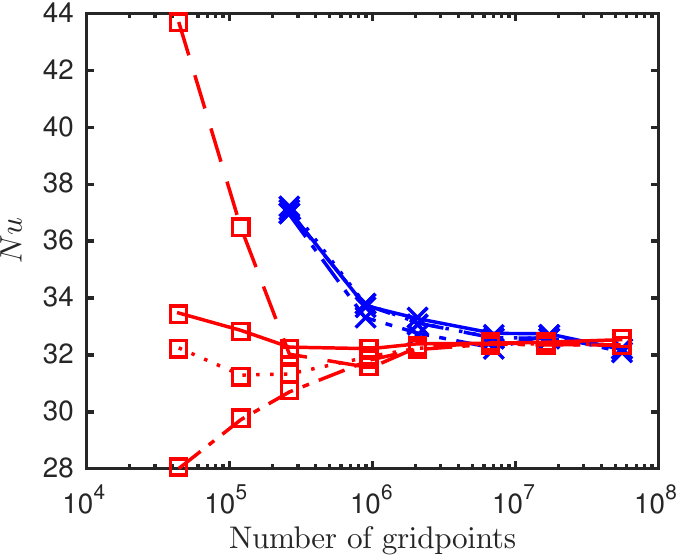}
\includegraphics[width=0.33\textwidth]{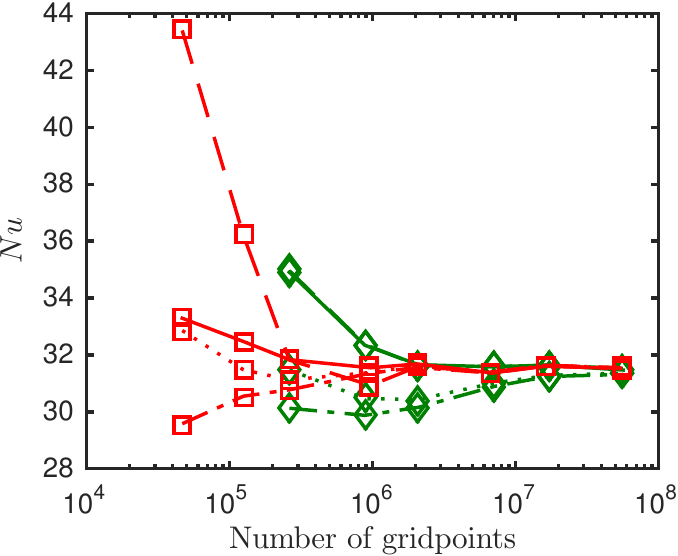}
\includegraphics[width=0.33\textwidth]{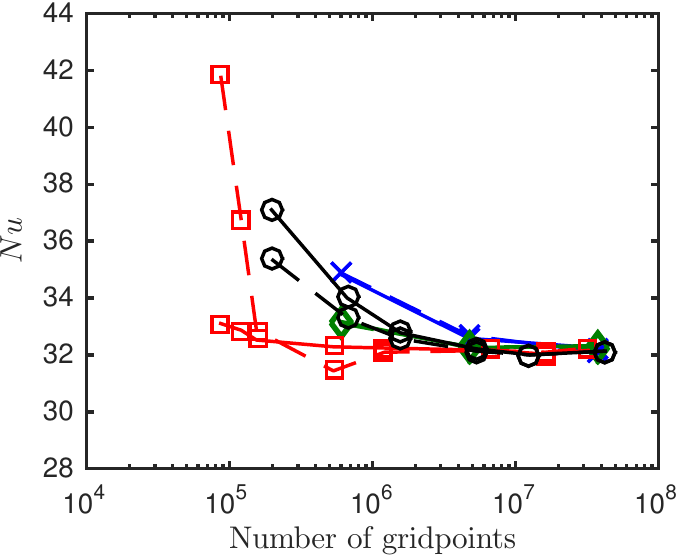}
\end{center}
\caption{$\Nus$ against the number of grid points for different geometries: (a) Periodic domain (b) Cubic domain (c) Cylindrical domain. Markers: $\color{blue}\times$ \textsc{AFID}, $\color{darkgreen}\diamond$ $\textsc{Goldfish}$, $\color{red}\square$ \textsc{Nek5000}, $\circ$ \textsc{OpenFOAM}. Lines: {-} $\Nus_{vol}$, $- - -$ $\Nus_{pl}$, $- \cdot -$ $\Nus_{kin}$, $\cdots$ $\Nus_{th}$.}
\label{fig:nu_vs_dx}
\end{figure*}

\subsection{Convergence test at $\Ray = 10^8$}
\label{sec:convergence}
\noindent Here we present a classical convergence test in which a number of grid refinements is undertaken to assess the sensitivity of the results to spatial resolution. Because of the fundamental differences between the spatial discretization techniques, the simulations on the different levels of grid refinement need to be performed with slightly different meshes. For a fair comparison, at any grid refinement level, the total number of grid points or degrees of freedom (which for simplicity we also call grid points) is kept similar for all codes. The mesh refinement is undertaken near the plates. The boundary layer thicknesses are estimated a priori using the scaling theory by Grossmann and Lohse (GL theory) \cite{gro00,shi10,ste10,ste13}. In the studied case of $\Pra=1$, the thicknesses of the kinetic and thermal boundary layers are similar. At any considered level of grid refinement, the total number of grid points, $N$, and the number of grid points inside the boundary layers, $N_{BL}$, of the chosen meshes are similar for all codes. The spatial resolutions and the corresponding values of $N$ and $N_{BL}$ are listed in the Appendix.

The $\Nus$ numbers are averaged over $300$ dimensionless time units after the solution approaches a statistically stationary state, which takes about $200$ time units, depending on the initial conditions. $\Nus$, obtained at different levels of grid refinement, versus the number of grid points, is presented in Fig.\ \ref{fig:nu_vs_dx}. We observe that at high grid resolutions all codes converge to the same result within a small time averaging error of about $0.5\%$. Different ways to calculate Nu and different codes lead to different convergence of the obtained $\Nus$ with increasing grid resolution. For example, the results for \textsc{Nek5000} at very coarse resolutions are quite inaccurate, but they converge quickly to the final value when the resolution is increased. We can interpret the results in Fig.\ \ref{fig:nu_vs_dx} as a good indication of convergence to nearly grid independent results, achieved by all codes, albeit at different spatial resolutions. Ultimate convergence assessment is hampered by the degree of time averaging uncertainty that remains, we come back to this momentarily.

\begin{table}[!t]
\centering
\caption{$\Nus$ obtained with the highest spatial resolutions for the periodic domain, where $N_x \times N_y \times N_z = 384 \times 384 \times 384$ for \textsc{AFID} and $N_x \times N_y \times N_z = 379 \times 379 \times 379$ for \textsc{Nek5000}.}
\label{tab:nu_peri}
\begin{tabular}{c c c c}
\toprule
 & $\Nus_{lo}$ & $\Nus_{up}$ & $\Nus_{vol}$ \\
\midrule
\textsc{AFID} & 32.24 & 32.27 & 32.18 \\
\textsc{Nek5000} & 32.29 & 32.41 & 32.54 \\
\midrule
Average & \multicolumn{3}{c}{$32.32 \pm 0.24\:(0.73\%)$} \\
\bottomrule
\end{tabular}
\end{table}

\begin{table}[!t]
\centering
\caption{$\Nus$ obtained with the highest spatial resolutions for the cubic container, where $N_x \times N_y \times N_z = 384 \times 384 \times 384$ for $\textsc{Goldfish}$ and $N_x \times N_y \times N_z = 379 \times 379 \times 379$ for \textsc{Nek5000}.}
\label{tab:nu_cube}
\begin{tabular}{c c c c}
\toprule
 & $\Nus_{lo}$ & $\Nus_{up}$ & $\Nus_{vol}$ \\
\midrule
\textsc{Goldfish} & 31.56 & 31.49 & 31.47 \\
\textsc{Nek5000} & 31.53 & 31.58 & 31.53 \\
\midrule
Average & \multicolumn{3}{c}{$31.53 \pm 0.08\:(0.24\%)$} \\
\bottomrule
\end{tabular}
\end{table}

\begin{table}[!t]
\centering
\caption{$\Nus$ obtained with the highest spatial resolutions for the cylindrical container, where $N_r \times N_\phi \times N_z = 192 \times 512 \times 384$ for \textsc{RBflow}/$\textsc{Goldfish}$, $N_{xy} \times N_z = 85009 \times 384$ for \textsc{Nek5000}, and $N_{xy} \times N_z = 110592 \times 384$ for \textsc{OpenFOAM}.}
\label{tab:nu_cyl}
\begin{tabular}{c c c c}
\toprule
 & $\Nus_{lo}$ & $\Nus_{up}$ & $\Nus_{vol}$ \\
\midrule
\textsc{RBflow} & 32.08 & 32.15 & 32.24 \\
\textsc{Goldfish} & 32.19 & 32.31 & 32.33 \\
\textsc{Nek5000} & 32.26 & 32.23 & 32.16 \\
\textsc{OpenFOAM} & 32.16 & 32.10 & 32.13 \\
\midrule
Average & \multicolumn{3}{c}{$32.20 \pm 0.14\:(0.44\%)$} \\
\bottomrule\end{tabular}
\end{table}

\begin{figure*}[!t]
\begin{center}
\includegraphics[width=0.33\textwidth]{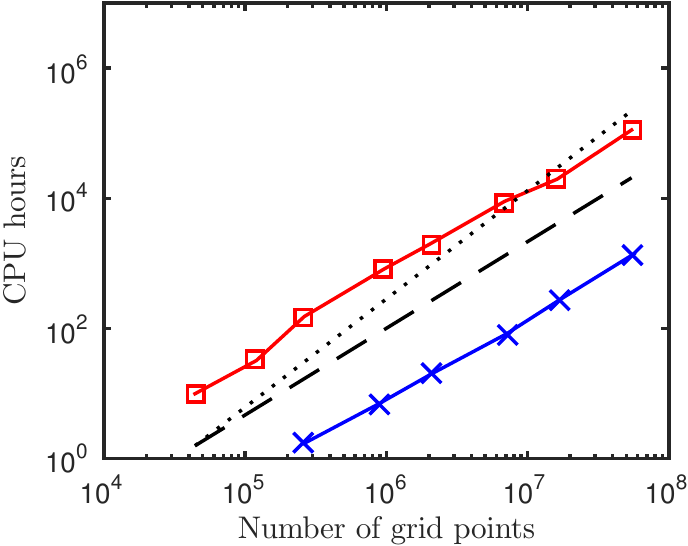}
\includegraphics[width=0.33\textwidth]{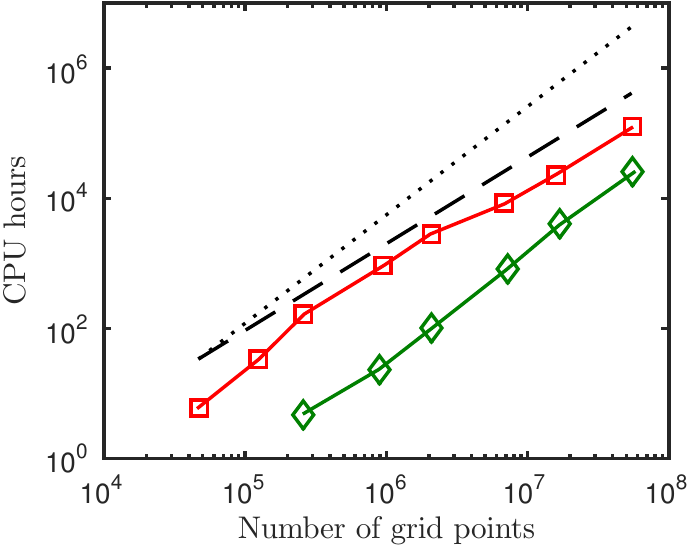}
\includegraphics[width=0.33\textwidth]{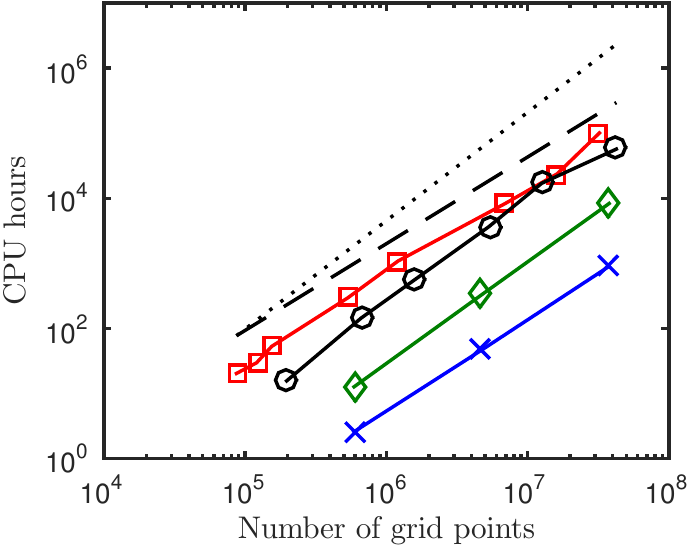}
\end{center}
\caption{Computational cost against the number of grid points for different geometries: (a) Periodic domain (b) Cubic domain (c) Cylindrical domain. $\color{blue}\times$ \textsc{AFID}/\textsc{RBflow}, $\color{darkgreen}\diamond$, \textsc{Goldfish}, $\color{red}\square$ \textsc{Nek5000}, $\circ$ \textsc{OpenFOAM}, $- - -$ $O(N^{4/3})$ and $\cdots$ $O(N^{5/3})$.}
\label{fig:cpu_cost}
\end{figure*}

For a fixed statistical averaging interval, the costs are proportional to the mesh size, both in space and time. For this cost-estimate to hold, it is required that the iterative solvers converge in a number of steps that is approximately constant during the time-interval over which the statistical averaging is performed. With suitable preconditioners such can be realized, as was observed for the corresponding codes. The Courant-Friedrichs-Lewy (CFL) condition and the numerical stability of the simulations were the two restrictions on the time stepping that were taken into account in all conducted simulations. As one can see in Fig.\ \ref{fig:cpu_cost}, the computational costs scale from $\mathbb{O}(N^{4/3})$ to $\mathbb{O}(N^{5/3})$ in all cases, in a full accordance to the von Neumann stability of the schemes used in the corresponding codes. For the schemes, which are optimal with respect to the von Neumann stability, the time step size $\tau$ is taken proportional to the mesh width $h$, which in turn is proportional to $N^{-1/3}$. This leads to the scaling of the computational costs with the mesh size as $\mathbb{O}(N^{4/3})$. Apart from the von Neumann stability, there exists also another restriction on the time stepping in accurate DNS, which is the resolution of the Kolmogorov time micro-scales; we will come back to this issue in section \ref{sec:conclusion}.

Further, in Fig.\ \ref{fig:cpu_cost}, we observe that the general purpose codes designed for unstructured grids in complex geometries are more expensive than those for structured grids, exploiting the periodic directions of the geometry or, at least the possibility to separate variables due to the regularity of the grids. For example, \textsc{AFID} and \textsc{OpenFOAM} are both second-order accurate, but \textsc{OpenFOAM} is much closer to \textsc{Nek5000} in terms of computational cost. Apart from the grid organization, the order of the schemes, used in the codes, influences the computational load. Obviously, the higher-order schemes need more operations per time step than the lower-order schemes. Also the way to solve the Poisson equation for the pressure-like function determines the efficiency. In general purpose codes only iterative computationally intensive solvers can be employed, while specialized codes can use direct solvers, which are much more efficient. Also the parallel scalability of the codes influences the total computational costs. \textsc{OpenFOAM}, for example, is characterized by quite modest scalability, compared to the other considered codes~\cite{axt16,riv11}. In our simulations with \textsc{OpenFOAM}, we used a sufficiently low number of cores, such that we operate only in the range of good parallel efficiency. In that case, the measurements of computational time are not affected significantly by possible effects of non-ideal scalability.

\begin{figure*}[!t]
\begin{center}
\includegraphics[width=0.33\textwidth]{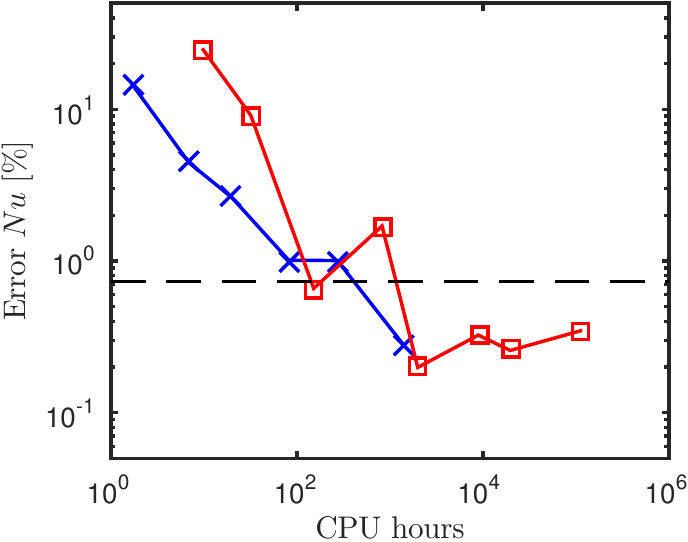}
\includegraphics[width=0.33\textwidth]{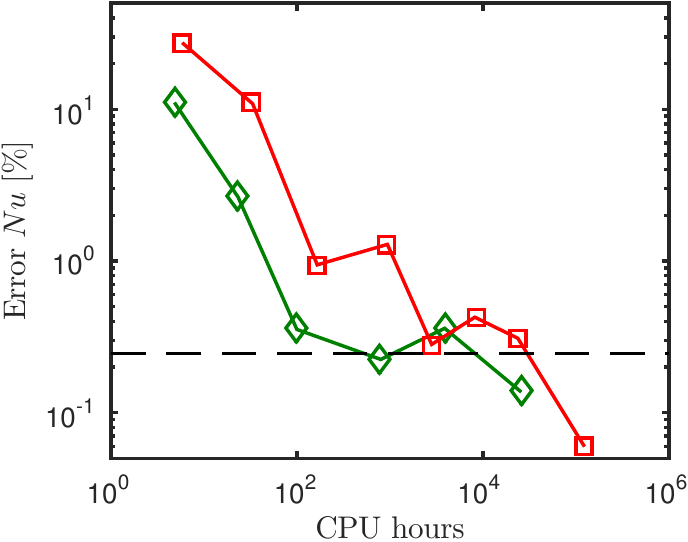}
\includegraphics[width=0.33\textwidth]{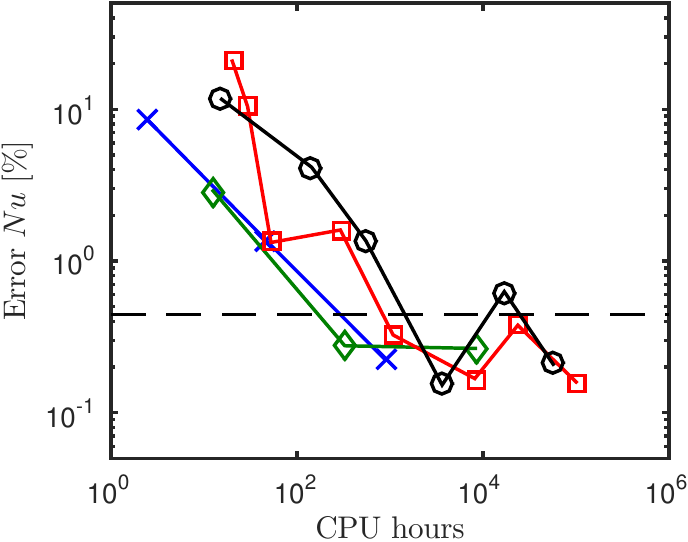}
\end{center}
\caption{Average error of $\Nus$ against the computational cost for different geometries: (a) Periodic domain (b) Cubic domain (c) Cylindrical domain. $\color{blue}\times$ \textsc{AFID}/\textsc{RBflow}, $\color{darkgreen}\diamond$ \textsc{Goldfish}, $\color{red}\square$ \textsc{Nek5000}, $\circ$ \textsc{OpenFOAM}. The dashed line indicates the $95\%$ confidence level of the reference value given in Tables \ref{tab:nu_peri}, \ref{tab:nu_cube}, and \ref{tab:nu_cyl}.}
\label{fig:error_vs_cost}
\end{figure*}

The $\Nus$ values obtained in the highest resolution simulations are listed in Tables \ref{tab:nu_peri}, \ref{tab:nu_cube}, and \ref{tab:nu_cyl}, for the periodic domain, cube, and cylinder, respectively. These data can be used to further quantify the error in $\Nus$ versus computational costs. To arrive at this we undertake a number of steps. First, we calculate the average $\Nus$ as reference points. We calculate the standard deviation from the data given in the tables and use twice the standard deviation as a $95\%$ confidence interval for the average values. These average $\Nus$ are subsequently used as a reference value to calculate the error in $\Nus_{lo}$, $\Nus_{up}$, and $\Nus_{vol}$ separately for the different codes and grids. After that, we take the average of those individual errors, and show the average error against the computational cost in Fig.\ \ref{fig:error_vs_cost}. The error decreases with increasing cost, until it becomes comparable to the confidence bounds. At that point, the error is dominated by the time averaging error, and no longer due to the spatial discretization. Since the computational costs for simulations of $300$ dimensionless time units are already considerably large we did not pursue a further reduction in the time averaging error. In fact, such time averaging error will tend to zero at a rate inversely proportional to the square root of the simulation time. Hence, only at extreme costs one could perceive a significant reduction of the time averaging error. Such resources are not available for this study and are also not required to establish the main conclusions.

In the periodic domain, \textsc{AFID} was found to be much faster than \textsc{Nek5000}. At a given computational cost, a much higher number of grid points can be afforded with \textsc{AFID}. On the other hand, this significant difference in computational cost between \textsc{Nek5000} and \textsc{AFID} when counting the number of grid points only, is considerably reduced when counting the actually achieved level of precision of the $\Nus$ prediction. Clearly, the higher-order method used in \textsc{Nek5000} is beneficial at reducing the gap with \textsc{AFID} in error versus cost considerations. This is illustrated concisely in Fig.\ \ref{fig:error_vs_cost}. For the cubic container we observe a similar situation: the specialized code \textsc{Goldfish} is much faster than \textsc{Nek5000}, see Fig.~\ref{fig:cpu_cost}b. And again, the efficiency of \textsc{Nek5000} becomes closer to that by \textsc{Goldfish} when the convergence of $\Nus$ is taken into account. When the error of the $\Nus$ calculation is above the confidence bound, \textsc{Goldfish} calculates $\Nus$ up to tenfold more accurately than \textsc{Nek5000}, for given computational costs. In the cylindrical container, \textsc{RBflow} and \textsc{Goldfish} are up to a factor ten faster than \textsc{OpenFOAM} for a given level of accuracy, while \textsc{RBflow} and \textsc{OpenFOAM} are both of second order. This illustrates the penalty that comes with the use of a general purpose code compared to a dedicated specialized code. \textsc{Nek5000} falls roughly in between \textsc{RBflow} and \textsc{OpenFOAM}. Overall, Fig.\ \ref{fig:error_vs_cost} shows that the large differences in speed that appear in Fig.\ \ref{fig:cpu_cost} (with \textsc{AFID}/\textsc{RBflow} the most efficient when it comes to costs of simulation with a certain number of grid points) decrease when counting the error in $\Nus$ versus computational costs due to the usage of higher-order schemes in the other codes.
 
In Fig.~\ref{fig:error_vs_cost} one can also see that the accuracy of the general purpose codes behave non-monotonically with increasing computational time. The strong oscillations in the behavior of the $\Nus$-error versus the costs are explained not only by the restricted time of statistical averaging, but mainly by the usage of the iterative solvers within these codes. While the dedicated codes \textsc{AFID/RBflow} and \textsc{Goldfish} use direct solvers and get the corresponding solutions at the machine accuracy, the iterative solvers of the general purpose codes stop iterations with a certain residual error. Furthermore, the used iterative solvers generally do not guarantee a monotonic reduction of the errors with increasing number of conducted iterations. This holds in particular for the generalized minimal residual method used in \textsc{Nek5000}. This makes prediction of the accuracy versus computation costs for general purpose codes less trivial and somewhat uncertain.

\subsection{Robustness against under-resolution}
\label{sec:high_ra}
\begin{figure}[!t]
\centering
\includegraphics[width=0.40\textwidth]{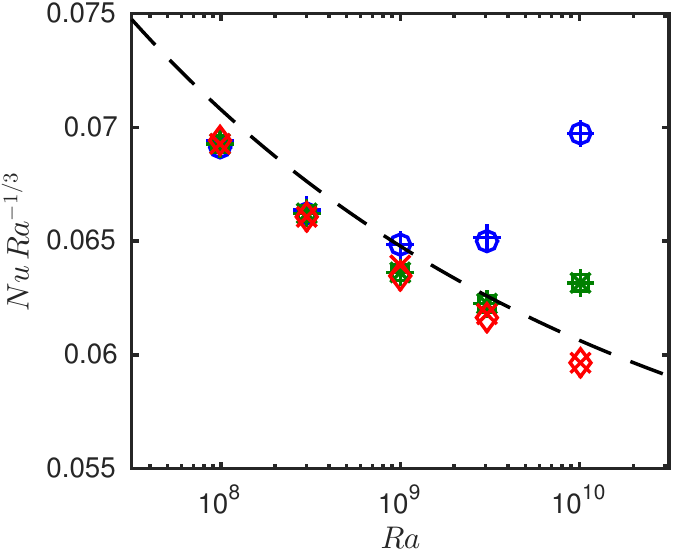}
\caption{$\Nus$ versus $\Ray$ in cylindrical container of $\Gamma=1$. These results of deliberately under-resolved DNS are obtained at a fixed computational mesh, which is too coarse for large $\Ray$. $\color{blue} +$ $\Nus_{vol}$ (\textsc{RBflow}), $\color{blue} \circ$ $\Nus_{pl}$ (\textsc{RBflow}), $\color{darkgreen} \ast$ $\Nus_{vol}$ (\textsc{Goldfish}), $\color{darkgreen} \square$ $\Nus_{pl}$ (\textsc{\textsc{Goldfish}}), $\color{red} \times$ $\Nus_{vol}$ (\textsc{Nek5000}), $\color{red} \diamond$ $\Nus_{pl}$ (\textsc{Nek5000}), $- - -$ Grossmann-Lohse theory \cite{ste13}. Note that a correct $\Nus$ does not imply a well-resolved flow, see Fig.~\ref{fig:snapshotsNek5000}
}
\label{fig:nu_vs_ra}
\end{figure}

\begin{figure*}[!t]
\begin{centering}
\includegraphics[width=0.99\textwidth]{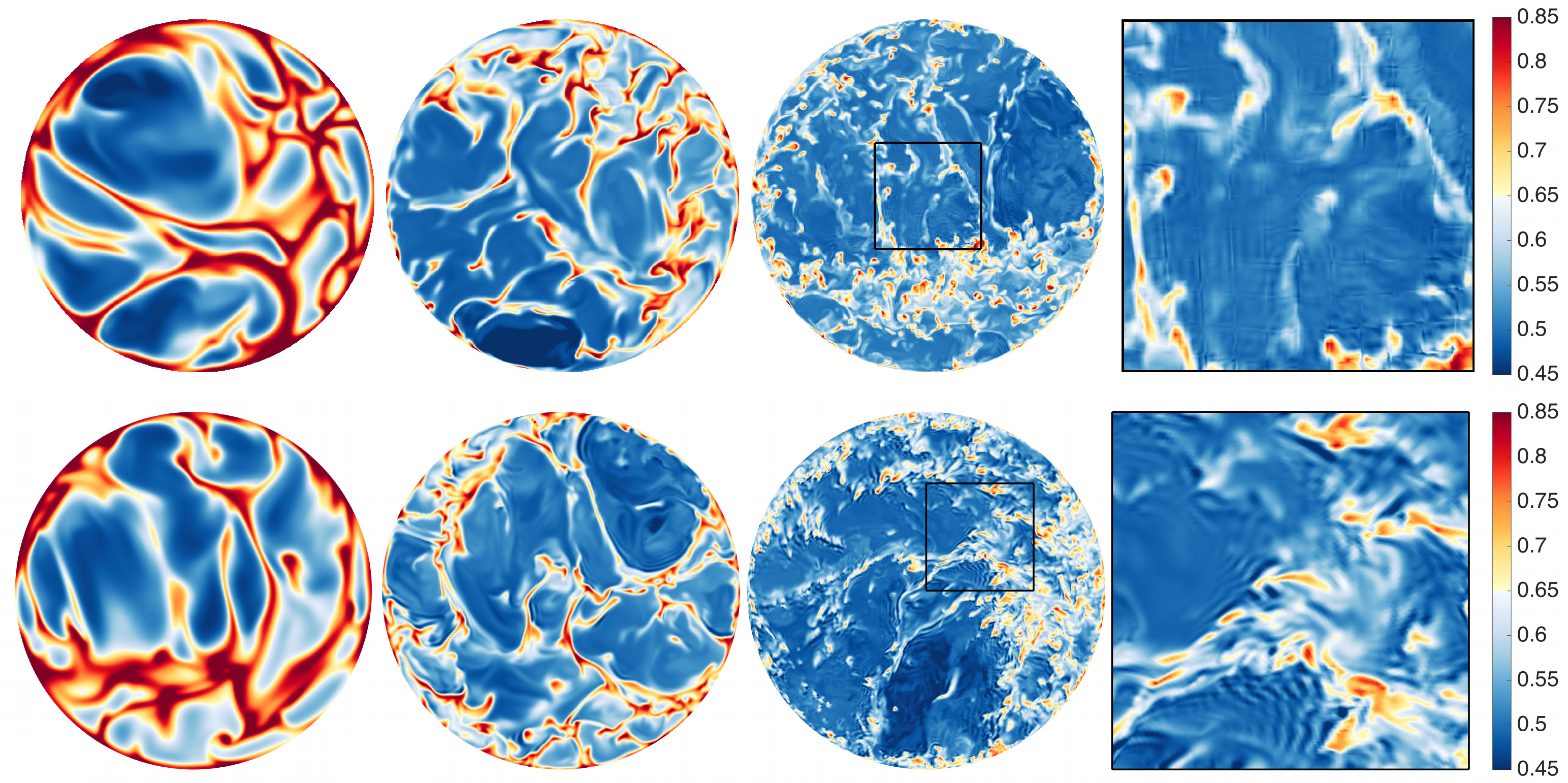}
\end{centering}
\caption{Temperature field at $z/H=0.0152$ from the \textsc{Nek5000} code (top row) and \textsc{RBflow} (bottom row) at (from left to right) $\Ray=10^8$, $\Ray=10^9$ and $\Ray=10^{10}$. For the latter two $\Ray$ values the chosen grid resolution is insufficient. Note that in the \textsc{Nek5000} snapshots the imprint of the computational grid is clearly visible in the higher $\Ray$ number cases, even though the $\Nus$ number from the simulations looks reasonable, see Fig.\ \ref{fig:nu_vs_ra}. The ripples in the \textsc{RBflow} snapshots are observed near sharp gradients when the resolution is insufficient.}
\label{fig:snapshotsNek5000}
\end{figure*}

\noindent At $\Ray = 10^8$, we are able to compute an accurate reference solution that is converged with respect to the spatial resolution independent of which code was adopted, (see Fig.\ \ref{fig:nu_vs_dx}). At higher $\Ray$ numbers, the computation of such a reference solution for all codes becomes too expensive, however. As an alternative, we can compare the codes in a different, somewhat more qualitative, way by increasing the $\Ray$ number while keeping the spatial resolution fixed. In this case, we use the meshes with the highest resolution adopted for $\Ray = 10^8$ in the cylindrical container. As $\Ray$ increases, the effect of insufficient resolution will unavoidably show up sooner or later, which indicates the robustness of the codes, against under-resolution. $\Nus$, compensated with $\Ray^{-1/3}$, is plotted against $\Ray$ in Fig.\ \ref{fig:nu_vs_ra}. Initially, the results of the three codes at $\Ray = 10^8$ are all very close to each other with the differences between the results less than $0.5\%$. The values of $\Nus$ are also very close to the prediction by the GL theory with the deviations between the simulation results and the GL predictions less than 1\%. As $\Ray$ increases the different robustnesses of the various codes against deliberate under-resolution become apparent. \textsc{Nek5000} shows the smallest deviation from the theoretical scaling of $\Nus$, and \textsc{RBflow} the largest. However, we emphasize that this robustness of $\Nus$ for \textsc{Nek5000} against deliberate underresolution does not imply that other flow features would still be well represented. E.g., in Fig.\ \ref{fig:snapshotsNek5000} we show temperature snapshots for well resolved and deliberately underresolved simulations with \textsc{Nek5000}. The latter clearly show a pronounced imprint of the computational grid at $Ra = 10^{10}$ even though the $Nu$ number predictions are not affected much. \textsc{Nek5000} does not show the imprint of the mesh at $Ra = 10^9$, but the effect of underresolution can be seen in very subtle ripples near high gradients. Similar plots obtained from \textsc{RBflow} also show an inadequacy of the grid resolution at $Ra=10^9$. Underresolution appears to affect \textsc{Nek5000} predictions somewhat less than is seen for \textsc{RBflow}.

This comparison clearly indicates that the agreement of $\Nus$ with the theoretical prediction (and among the values obtained from the various definitions) is not enough to assess the adequacy of the spatial resolution of the numerical simulation. Additional quantities have to be analyzed, such as the instantaneous temperature snapshots or rms profiles, in order to clarify this issue.

\section{Conclusions and outlook}
\label{sec:conclusion}
\noindent In this paper, we have compared several codes for the simulation of turbulent RB convection in a number of typical geometries. Particular attention has been given to the heat transport in the turbulent flow, which is quantified by $\Nus$. The computational efficiency of the codes is determined with reference to fully converged simulations at relatively high spatial resolutions. We observed significant differences between the codes in terms of computational costs, i.e.\ the specialized \textsc{AFID}/\textsc{RBflow} and \textsc{Goldfish} code clearly outperform \textsc{Nek5000} and \textsc{OpenFOAM}. Thus, we note that a considerable saving in computational costs can be achieved by employing an optimized code for a simple geometry compared to general purpose codes designed for complex geometries. The benefit of general purpose codes like \textsc{Nek5000} and \textsc{OpenFOAM} is of course that they are much wider applicable than specialized codes, which need to be specifically tuned per case. 

The usage of unstructured grids in the general purpose codes requires iterative solutions of the governing equations on each time step. This leads to higher computational costs and makes these codes less predictable with respect to the accuracy of the calculation of $\Nus$ with growing computational costs. Also the scalability of the \textsc{OpenFOAM} codes on supercomputers leaves much to be desired. All this leads to the fact that \textsc{AFID}/\textsc{RBflow}, being also the second-order as \textsc{OpenFOAM} in the considered configuration, is at least ten times faster than \textsc{OpenFOAM}, while providing the same level of accuracy. Therefore we conclude that \textsc{OpenFOAM}, at least in the analyzed configuration, which is the most popular in engineering, is not optimal for scientific investigations of high $\Ray$ number thermal convection.

Also, among the other codes, \textsc{AFID}/\textsc{RBflow} is clearly the fastest one. It is up to tenfold faster than \textsc{Goldfish} and up to hundredfold faster than \textsc{Nek5000}. However, when the accuracy of the Nu calculation is taken into account, the efficiency of \textsc{Goldfish} and \textsc{AFID}/\textsc{RBFlow} is similar, while \textsc{Nek5000} and \textsc{OpenFOAM} are up to $10$ times slower. When in a certain numerical study the point of interest is an integral quantity (zero moment), like $\Nus$ or Reynolds number, or when the profiles of the mean temperature or velocity are aimed to be studied (first moments), the advantages of the usage of the second-order code \textsc{AFID}/\textsc{RBflow} are clearly pronounced. It is extremely fast and calculates these quantities precisely on sufficiently fine meshes. This is partly thanks to the implementation of \textsc{AFID}/\textsc{RBflow}, which is highly optimized, and scales excellently on large number of cores \cite{zhu17}.

Finally, we give a general estimate of the complexity of the DNS of turbulent RB convection in the classical regime and in the ultimate regime, which is to be studied in the future. As we already mentioned in section \ref{sec:convergence}, apart from the CFL-condition and the von Neumann stability, there exists also another restriction on the time stepping in accurate DNS, which is the resolution of the Kolmogorov time microscales. Note that the Kolmogorov microscale in space, $\eta\equiv(\mu^3/\avg{\varepsilon}_V)^{1/4}$, and the microscale in time, $\eta_\tau\equiv(\mu/\avg{\varepsilon}_V)^{1/2}$, are related as $\mu\,\eta_\tau\sim\eta^2$ with $\mu\equiv\sqrt{\Pr/\Ray}$. Thus, the optimal (not over-resolved but accurate) DNS, which resolve both, the Kolmogorov time microscale $\eta_\tau$ and the Kolmogorov spatial microscales $\eta$, will lead to the scaling of the computational costs with the grid size $N$ at least as $\mathbb{O}(\mu N^{5/3})$. Since $\avg{\varepsilon}_V= (\Nus-1)/\sqrt{\Pra \Ray}$, for a fixed $\Pra$, the computational costs in accurate DNS must grow at least as $\mathbb{O}(\Nus^{5/4}\,\Ray^{3/4})$. Therefore, in the classical regime, where $\Nus\sim\Ray^{1/3}$, the cost will increase with $\Ray$ at least as $\mathbb{O}(\Ray^{7/6})$, while for the ultimate regime, where the scaling $\Nus\sim\Ray^{1/2}$ is expected, the anticipated computational costs in accurate DNS are at least $\mathbb{O}(\Ray^{11/8})$.

Before concluding this paper we wish to point once more out that comparing $\Nus$ obtained by the numerical simulation with the expected value is not a reliable criterion to assess its validity. In fact, we showed that deliberately under-resolved simulations performed with higher order codes show a small error in $\Nus$ while producing temperature fields with strong unphysical oscillations. Instantaneous snapshots of temperature and profiles of higher order moments have to be evaluated, together with $\Nus$, in order to establish the quality of a numerical simulation.

\section*{Acknowledgements}
\noindent 
GLK is funded by Foundation for Fundamental Research on Matter (FOM), part of the Netherlands Organization for Scientific Research (NWO). DL and RV are funded by the Netherlands Center for Multiscale Catalytic Energy Conversion (MCEC), an NWO Gravitation program funded by the Ministry of Education, Culture and Science of the government of the Netherlands. OS and SH are funded by the Deutsche Forschungsgemeinschaft (DFG) under the grants Sh~405/4-2 and Ho~5890/1-1, respectively. DL and OS thank DFG Priority Programme SPP 1881 "Turbulent Superstructures". We thank NWO for granting us computational time on Cartesius cluster from the Dutch Supercomputing Consortium SURFsara under grants SH-061 and SH-015 and also are grateful to the Leibniz Rechenzentrum (LRZ) for providing us computational resources under the grant pr84pu.

\onecolumn

\clearpage
\section*{Appendix: Spatial resolution and Nusselt number at $\Ray = 10^8$ and $\Gamma=1$.} \label{sec:resolution}
\FloatBarrier

\begin{table}[!h]
\centering
\caption{Spatial resolution and $\Nus$ numbers for \textsc{AFID} in the periodic domain.}
\begin{tabular}{r r r r r r r}
\toprule
$N$ & $N_{BL}$ & $\Nus_{lo}$ & $\Nus_{up}$ & $\Nus_{vol}$ & $\Nus_{kin}$ & $\Nus_{th}$ \\
\midrule
$64^3$ & 3 & 37.02 & 37.00 & 37.06 & 36.94 & 37.24\\
$96^3$ & 4 & 33.73 & 33.81 & 33.74 & 33.32 & 33.80\\
$128^3$ & 5 & 33.11 & 33.12 & 33.27 & 32.76 & 33.12\\
$192^3$ & 8 & 32.57 & 32.62 & 32.76 & 32.24 & 32.60\\
$256^3$ & 11 & 32.57 & 32.62 & 32.75 & 32.69 & 32.59\\
$384^3$ & 16 & 32.24 & 32.27 & 32.18 & 32.10 & 32.25\\
\bottomrule
\end{tabular}
\label{tab:nu_AFID_periodic}
\end{table}

\begin{table}[!h]
\centering
\caption{Spatial resolution and $\Nus$ numbers for \textsc{Nek5000} in the periodic domain.}
\begin{tabular}{r r r r r r r r}
\toprule
$E$ & $N$ & $N_{BL}$ & $\Nus_{lo}$ & $\Nus_{up}$ & $\Nus_{vol}$ & $\Nus_{kin}$ & $\Nus_{th}$ \\
\midrule
$5^3$ & $36^3$ & 2 & 43.78 & 43.63 & 33.47 & 28.01 & 32.25\\
$7^3$ & $50^3$ & 2 & 36.47 & 36.48 & 32.86 & 29.72 & 31.29\\
$9^3$ & $64^3$ & 3 & 32.01 & 32.04 & 32.27 & 30.70 & 31.33\\
$14^3$ & $99^3$ & 4 & 31.57 & 31.53 & 32.22 & 31.83 & 31.95\\
$18^3$ & $127^3$ & 5 & 32.30 & 32.22 & 32.39 & 32.19 & 32.23\\
$27^3$ & $190^3$ & 8 & 32.41 & 32.46 & 32.41 & 32.38 & 32.41\\
$36^3$ & $253^3$ & 11 & 32.31 & 32.43 & 32.45 & 32.35 & 32.38\\
$54^3$ & $379^3$ & 16 & 32.29 & 32.41 & 32.54 & 32.35 & 32.37\\
\bottomrule
\end{tabular}
\label{tab:nu_Nek5000_periodic}
\end{table}

\begin{table}[!h]
\centering
\caption{Spatial resolution and $\Nus$ numbers for $\textsc{Goldfish}$ in the cubic container.}
\begin{tabular}{r r r r r r r}
\toprule
$N$ & $N_{BL}$ & $\Nus_{lo}$ & $\Nus_{up}$ & $\Nus_{vol}$ & $\Nus_{kin}$ & $\Nus_{th}$ \\
\midrule
$64^3$ & 3 & 34.93 & 34.99 & 34.91 & 30.13 & 31.47\\
$96^3$ & 4 & 32.38 & 32.36 & 32.34 & 29.88 & 30.46\\
$128^3$ & 5 & 31.56 & 31.70 & 31.66 & 30.16 & 30.44\\
$192^3$ & 8 & 31.67 & 31.51 & 31.58 & 30.89 & 31.01\\
$256^3$ & 11 & 31.65 & 31.63 & 31.64 & 31.24 & 31.28\\
$384^3$ & 16 & 31.56 & 31.49 & 31.47 & 31.31 & 31.36\\
\bottomrule
\end{tabular}
\end{table}

\begin{table}[!h]
\centering
\caption{Spatial resolution and $\Nus$ numbers for \textsc{Nek5000} in the cubic container.}
\begin{tabular}{r r r r r r r r}
\toprule
$E$ & $N$ & $N_{BL}$ & $\Nus_{lo}$ & $\Nus_{up}$ & $\Nus_{vol}$ & $\Nus_{kin}$ & $\Nus_{th}$ \\
\midrule
$5^3$ & $36^3$ & 2 & 43.44 & 43.38 & 33.29 & 29.58 & 32.86\\
$7^3$ & $50^3$ & 2 & 36.13 & 36.28 & 32.47 & 30.55 & 31.48\\
$9^3$ & $64^3$ & 3 & 31.76 & 31.88 & 31.83 & 30.77 & 31.12\\
$14^3$ & $99^3$ & 4 & 30.93 & 30.93 & 31.55 & 31.36 & 31.38\\
$18^3$ & $127^3$ & 5 & 31.53 & 31.64 & 31.68 & 31.56 & 31.62\\
$27^3$ & $190^3$ & 8 & 31.40 & 31.41 & 31.37 & 31.36 & 31.38\\
$36^3$ & $253^3$ & 11 & 31.56 & 31.71 & 31.60 & 31.62 & 31.63\\
$54^3$ & $379^3$ & 16 & 31.53 & 31.58 & 31.53 & 31.52 & 31.55\\
\bottomrule
\end{tabular}
\end{table}

\begin{table}[!h]
\centering
\caption{Spatial resolution and $\Nus$ numbers for \textsc{RBflow} in the cylindrical container.}
\begin{tabular}{r r r r r r r r r}
\toprule
$N_r$ & $N_\phi$ & $N_z$ & $N_{BL}$ & $\Nus_{lo}$ & $\Nus_{up}$ & $\Nus_{vol}$ \\
\midrule
48 & 128 & 96 & 4 & 34.95 & 34.86 & 34.85\\
96 & 256 & 192 & 8 & 32.59 & 32.76 & 32.58\\
192 & 512 & 384 & 16 & 32.08 & 32.15 & 32.24\\
\bottomrule
\end{tabular}
\end{table}

\begin{table}[!h]
\centering
\caption{Spatial resolution and $\Nus$ numbers for $\textsc{Goldfish}$ in the cylindrical container.}
\begin{tabular}{r r r r r r r r r}
\toprule
$N_r$ & $N_\phi$ & $N_z$ & $N_{BL}$ & $\Nus_{lo}$ & $\Nus_{up}$ & $\Nus_{vol}$ \\
\midrule
48 & 128 & 96 & 4 & 33.20 & 33.02 & 33.18\\
96 & 256 & 192 & 8 & 32.40 & 32.19 & 32.26\\
192 & 512 & 384 & 16 & 32.19 & 32.31 & 32.33\\
\bottomrule
\end{tabular}
\end{table}

\begin{table}[!h]
\centering
\caption{Spatial resolution and $\Nus$ numbers for \textsc{Nek5000} in the cylindrical container. $E_{xy}$ denotes the number of spectral elements in a horizontal cross section, and $E_z$ in the vertical direction.}
\begin{tabular}{r r r r r r r r r r}
\toprule
$E_{xy}$ & $E_z$ & $N_{xy}$ & $N_z$ & $N_{BL}$ & $\Nus_{lo}$ & $\Nus_{up}$ & $\Nus_{vol}$ \\
\midrule
48 & 5 & 2409 & 36 & 2 & 41.78 & 41.85 & 33.10\\
48 & 7 & 2409 & 50 & 2 & 36.91 & 36.62 & 32.87\\
48 & 9 & 2409 & 64 & 3 & 32.68 & 33.06 & 32.52\\
108 & 14 & 5377 & 99 & 4 & 31.47 & 31.41 & 32.29\\
192 & 18 & 9521 & 127 & 5 & 32.07 & 32.12 & 32.24\\
432 & 27 & 36100 & 190 & 8 & 32.14 & 32.18 & 32.16\\
768 & 36 & 64009 & 253 & 11 & 32.11 & 32.08 & 32.03\\
1728 & 54 & 85009 & 379 & 16 & 32.26 & 32.23 & 32.16\\
\bottomrule
\end{tabular}
\end{table}

\begin{table}[!h]
\centering
\caption{Spatial resolution and $\Nus$ numbers for \textsc{OpenFOAM} in the cylindrical container.}
\begin{tabular}{r r r r r r r r}
\toprule
$N_{xy}$ & $N_z$ & $N_{BL}$ & $\Nus_{lo}$ & $\Nus_{up}$ & $\Nus_{vol}$ \\
\midrule
3072 & 64 & 3 & 35.43 & 35.30 & 37.10\\
6912 & 96 & 4 & 33.33 & 33.27 & 34.00\\
12288 & 128 & 5 & 32.62 & 32.52 & 32.78\\
27648 & 192 & 8 & 32.17 & 32.09 & 32.21\\
49152 & 256 & 11 & 31.90 & 32.10 & 32.00\\
110592 & 384 & 16 & 32.16 & 32.10 & 32.13\\
\bottomrule
\end{tabular}
\end{table}

\end{document}